\documentclass[preprint,prb,amsmath,amssymb,showpacs]{revtex4}
\usepackage{graphicx}
\usepackage{epsfig}

\begin{document}
\title{Less than perfect quantum wavefunctions in momentum-space: How
$\phi(p)$ senses disturbances in the force}

\author{M. Belloni}
\email{mabelloni@davidson.edu}
\affiliation{Department of Physics, Davidson College, Davidson, North Carolina
28035 USA}

\author{R. W. Robinett}
\email{rick@phys.psu.edu}
\affiliation{Department of Physics, The Pennsylvania State
University, University Park, Pennsylvania 16802 USA}

\begin{abstract}
We develop a systematic approach to determine the large $|p|$ behavior of the momentum-space wavefunction, $\phi(p)$,
of a one-dimensional quantum system
for which the position-space wavefunction, $\psi(x)$,
has a discontinuous
derivative at any order. We find that if the $k$th derivative of the
potential energy function for the system has a discontinuity, there is a corresponding
discontinuity in $\psi^{(k+2)}(x)$ at the same point.
This discontinuity leads directly to a power-law tail in the momentum-space
wavefunction proportional to $1/p^{k+3}$.
A number of familiar pedagogical examples are examined in this context, leading to a general derivation of the result.
\end{abstract}



\maketitle

\section{Introduction}
\label{sec:intro}

Much of the content of traditional courses in quantum mechanics
consists of solving the time-independent Schr\"odinger equation
in position space and applying the appropriate boundary conditions
to find the energy eigenstates, $\psi_{n}(x)$.

Student understanding of the connections between the
potential energy function of the system and the
behavior of $\psi_{n}(x)$ is becoming an increasingly important part of
the undergraduate curriculum. In quantum mechanics, as in
architecture and industrial design, it is true that
``\ldots form follows function \ldots'' \cite{form_follows_function}
and the detailed behavior of $\psi_{n}(x)$
(both its magnitude and local variation) are strongly correlated
with the behavior of the potential energy, $V(x)$. Analyses
focusing on these connections can appear as early as modern physics
courses (at the level of sketching wavefunctions)
up through formal implementations of the idea via approaches
such as the WKB method. Many textbooks and articles
\cite{robinett_classical_quantum,robinett_momentum,yoder}
compare the quantum mechanical probability densities
in both position- and momentum-space, $|\psi_{n}(x)|^2$ and $|\phi_{n}(p)|^2$,
and their classical analogs.

Although some of the standard textbook level examples, such as the
harmonic oscillator, give solutions which are infinitely differentiable, several of the most familiar
model problems are based on potential energy functions that have discontinuities in some
derivative, are discontinuous themselves (such as the step potential), or are
singular (the $\delta$-function and infinite well.)

All one-dimensional potentials must give solutions
for which $\psi_{n}(x)$ is continuous and for which
at least the expectation value of $p^2$ (necessary for
the Schr\"{o}dinger equation) is well-defined. Higher-order derivatives
of $\psi_{n}(x)$ can be discontinuous (or singular),
implying that expectation
values of higher powers of $p$ are not defined and repeated application
of the differential momentum operator, $\hat{p}$, will
``find'' such discontinuities in position space.

In momentum space the expectation value of powers of momentum
is given by
\begin{equation}
\langle p^k\rangle = \!\int_{-\infty}^{+\infty}\, p^{k}\, |\phi_{n}(p)|^2\,dp,
\label{basic_momentum_space_definition}
\end{equation}
and questions related to the evaluation of such average values will
necessarily be tied to the large $|p|$ behavior of $\phi_{n}(p)$ and
whether it leads to convergent integrals.
A natural question is to what extent the continuity properties of
$V(x)$ are reflected, directly or indirectly, in the asymptotic behavior of $\phi(p)$, which is the subject of this paper.
In other words, can we tell from the form of $V(x)$
how $\phi_{n}(p)$ behaves for large values of $|p|$?

We will demonstrate, first through examples, and then
by a more formal derivation, that if $V(x)$ has a discontinuity in its $k$th derivative, $V^{(k)}(x)$
at $x=a$, then there is a corresponding discontinuity in
$\psi^{(k+2)}(x)$, with
\begin{equation}
\psi^{(k+2)}(a^{+}) - \psi^{(k+2)}(a^{-})
=
\frac{2m}{\hbar^2} \left[
V^{(k)}(a^{+})
-
V^{(k)}(a^{-})
\right] \psi(a) .
\label{first_quote}
\end{equation}
The leading term
in the large $|p|$ expansion of $\phi(p)$ is directly related
to this generalized ``kink'' and is given by
\begin{equation}
\phi(p)
=(-i)^{k+3}\,e^{-ipa/\hbar} \frac{[\psi^{(k+2)}(a^{+}) - \psi^{(k+2)}(a^{-})]}{\sqrt{2\pi \hbar}}
\left(\frac{\hbar}{p}\right)^{k+3},
\label{second_quote}
\end{equation}
so that expectation values through $\langle p^{2(k+2)} \rangle$
are well-defined. The use of Eq.~(\ref{first_quote})
in Eq.~(\ref{second_quote})
demonstrates how the form of $\phi(p)$ depends on $V^{(k)}$.
If the wavefunction vanishes at the cusp, $\psi(a) = 0$,
then the leading-order term will be ${\cal O}(\hbar/p)^{k+4}$.
In this language, singular potentials (such as the $\delta$-function
or any function involving an infinite wall/barrier) correspond to $k=-1$,
and discontinuous potentials such as the step potential have $k=0$.
The symmetric linear potential, defined by $V(x) \propto |x|$
(with a cusp at $x=0$), which is discussed in
Sec.~\ref{sec:symmetric_linear_potential}, corresponds to $k=1$.

We will follow a pedagogical approach, in many ways following the path we took
in exploring the question.
We first examine the connection between the singular nature of
$V(x)$ and the large $|p|$ behavior of $\phi(p)$ for two familiar textbook-level systems, the $\delta$-function potential
(in Sec.~\ref{sec:delta_function}) and the infinite well
(in Sec.~\ref{sec:infinite_well}.) In Sec.~\ref{sec:quantum_bouncer} we discuss the same issues for
a single infinite wall potential,
the quantum bouncer. We
then illustrate in Sec.~\ref{sec:hints} the simple intuitive example
that first led us to the systematic expansion of $\phi(p)$, leading
to a formal general solution in Sec.~\ref{sec:formal_solution}.
We include a final exemplary case in
Sec.~\ref{sec:symmetric_linear_potential},
which illustrates a special circumstance in which the general result
requires more careful interpretation. We then review
our results, conclusions, and suggest further avenues of study.

\section{Single $\delta$-function}
\label{sec:delta_function}

A simple model system with a singular potential
is the single attractive $\delta$-function potential, defined as
\begin{equation}
V_\delta(x;a) = -g\delta (x-a).
\end{equation}
We integrate the Schr\"odinger equation over the interval
$(a^{-},a^{+}) = (a-\epsilon,a+\epsilon)$ and find the discontinuity
condition on the wavefunction
\begin{equation}
\psi'(a^{+}) - \psi'(a^{-}) = -\frac{2mg}{\hbar^2} \psi(a),
\label{cusp_condition}
\end{equation}
which can be used to determine the energy eigenvalue condition.
There is a single bound state solution, given by
\begin{equation}
\psi_0(x) = \sqrt{K_0} \, e^{-K_0|x-a|},
\end{equation}
where $K_0 \equiv mg/\hbar^2$, with bound state energy
\begin{equation}
E_0 = -|E_0| = - \frac{\hbar^2 K^2_0}{2m} = - \frac{mg^2}{2\hbar^2} .
\end{equation}

The expectation value of the potential energy is given by
\begin{equation}
\langle \psi_0| V(x) |\psi_0 \rangle = - g|\psi_0(a)|^2 = - gK_0
= - \frac{\hbar^2 K^2_0}{m} = -2\left(\frac{\hbar^2K^2_0}{2m}\right),
\end{equation}
which implies that the average value of the kinetic energy is
\begin{equation}
\langle \psi_0| \hat{T} |\psi_0 \rangle =
E_0 - \langle \psi_0| V(x)|\psi_0 \rangle
= \frac{\hbar^2K^2_0}{2m} . \label{eqthis}
\end{equation}
Equation~\eqref{eqthis} can be confirmed by using either of two equivalent expressions
for the expectation value of the kinetic energy
\begin{equation}
\langle \hat{T} \rangle =
- \frac{\hbar^2}{2m} \!\int_{-\infty}^{+\infty} \psi^*(x) \frac{d^2 \psi(x)}{dx^2}\, dx
= \frac{\hbar^2}{2m}\!\int_{-\infty}^{+\infty}
\left|\frac{d \psi(x)}{dx}\right|^2 dx
\label{x_ke} ,
\end{equation}
where an integration by parts was used to obtain the second equality
from the first.
From the second form in Eq.~(\ref{x_ke}), which requires only the
first derivative of $\psi(x)$, we have $|\psi'_0(x)| = K_0 \psi_0(x)$, giving
\begin{equation}
\langle \psi_0 |\hat{T} |\psi_0\rangle = \frac{\hbar^2 K^2_0}{2m}.
\end{equation}
The first form in Eq.~(\ref{x_ke}) can also be used if we note
that
\begin{equation}
\frac{d^2 \psi_0(x)}{dx^2} =
K^2_0 \psi_0(x) - 2K_0\sqrt{K_0}\delta(x-a),
\end{equation}
where the important $\delta$-function contribution arises from differentiating
the discontinuous $\psi'(x)$ at $x=a$,
using the relation $\Theta'(x) = \delta(x)$
for the Heaviside function.
This result is significant because it confirms that further derivatives
of $\psi(x)$ are not well defined, so that expectation values
of higher powers of $\hat{p}$ are not calculable.

The corresponding momentum-space wavefunction is given by
\begin{equation}
\phi_0(p) = \frac{1}{\sqrt{2\pi \hbar}}
\int_{-\infty}^{+\infty} \psi_0(x)\, e^{-ipx/\hbar}\,dx
= \sqrt{\frac{2p_0}{\pi}}\,
\left(\frac{p_0}{p^2 + p_0^2}\right)\,e^{-ipa/\hbar},
\label{single_delta_momentum_space}
\end{equation}
where $p_0 \equiv \hbar K_0 = mg/\hbar$.
Standard integrals then give
\begin{equation}
\langle \psi_0 |p^2 |\psi_0\rangle = \int_{-\infty}^{+\infty} p^2
|\phi_0(p)|^2\,dp = p_0^2,
\end{equation}
giving $\langle T \rangle = +(\hbar K_0)^2/2m$ as expected.

The large $|p|$ behavior of the momentum-space wavefunction is given by
$|\phi(p)| \propto 1/p^2$ which implies that expectation values of powers
of $p$ higher than $2$ will not lead to convergent integrals. This behavior is the
first hint of the connections between the continuity behavior
of $\psi(x)$ and the large $|p|$ behavior of $\phi(p)$.

As an aside, we note that we need to calculate expectation values
only of even powers of $p$, because for stationary state solutions
of bound state systems, the energy eigenfunctions can be put into purely
real form, which implies that $|\phi(-p)|^2 = |\phi(+p)|^2$ and
expectation values of odd powers vanish. The most obvious example is that
$\langle \psi_n |\hat{p}|\psi_n \rangle = \langle \phi_n |p|\phi_n\rangle = 0$
for any such state, corresponding to the fact that the particle
is equally likely to be found moving to the right ($p = +|p|$)
or the left ($p = -|p|$).

An alternative derivation of $\phi(p)$ which also gives the form
in Eq.~(\ref{single_delta_momentum_space})
involves Fourier transforming the Schr\"odinger equation directly
into momentum space \cite{lieber}
by multiplying by $\exp(-ipx/\hbar)$ and integrating
over all space, which gives
\begin{equation}
\frac{p^2}{2m} \phi_0(p)
- \frac{g}{\sqrt{2\pi \hbar}}
\int_{-\infty}^{+\infty} \delta(x-a)\, \psi_0(x)\, e^{-ipx/\hbar}\,dx
= -|E_0|\phi_0(p),
\end{equation}
or
\begin{equation}
\phi_0(p) = \frac{2m}{\sqrt{2\pi \hbar}}
\frac{g\psi(a)}{(p^2 + 2m|E_0|)}\, e^{-ipa/\hbar} .
\label{momentum_space_version}
\end{equation}
This form is useful because it allows an easy generalization if
multiple $\delta$-function potentials are present.
If
\begin{equation}
V(x) = - \sum_{i=1}^{N}g_i\delta(x-a_i),
\end{equation}
we immediately have that
\begin{equation}
\phi(p) = \left[\frac{2m}{\sqrt{2\pi \hbar}}
\frac{1}{(p^2 + 2m|E|)}\right]
\sum_{i=1}^{N} g_i \psi(a_i) e^{-ipa_i/\hbar} .
\end{equation}
The presence of multiple singularities in the potential (and multiple cusps
in the wavefunction) allows for interference between the
$\exp(-ipa_i/\hbar)$ phases, but still gives the overall $1/p^2$ behavior for
large $|p|$.

Looking forward to our general result,
we examine the $|p| \rightarrow \infty$ behavior of
Eq.~(\ref{momentum_space_version}) in a slightly different way. If we use
the connection in Eq.~(\ref{cusp_condition}), we find that the
large $|p|$ limit can be written in the form
\begin{equation}
\phi(p)
\to
\frac{1}{\sqrt{2\pi\hbar}} \, [\psi'(a^{-}) - \psi'(a^{+})]\, e^{-ipa/\hbar}\,\frac{\hbar^2}{p^2}
\qquad
(|p| \rightarrow \infty).
\label{first_general_result}
\end{equation}
This result is important because it shows that the large $p$ momentum-space
wavefunction can be written in a way that depends only on the properties
of the cusp in the wavefunction at $x=a$. 

Although derived for this case, we will find
that the result in Eqn.~(\ref{first_general_result})
 is the first non-trivial term is a systematic expansion of the
leading-order large $|p|$ behavior of $\phi(p)$ for
which the corresponding $\psi(x)$ has a discontinuity in some derivative.

\section{Infinite well}
\label{sec:infinite_well}
For an infinite well potential defined in the region $(0,L)$, the
normalized wavefunctions are
\begin{equation}
\psi_n(x) =
\begin{cases}
\sqrt{\dfrac{2}{L}}
\, \sin \left(\dfrac{n\pi x}{L}\right)
& (0 \leq x \leq L) \\
0 & (\mbox{$x<0$ or $L<x$)},
\end{cases}
\end{equation}
with $n=1,2,3,\ldots$. The spatial derivatives necessary to evaluate the kinetic energy
using either form in Eq.~(\ref{x_ke}) are given by
\begin{equation}
\frac{d\psi_n(x)}{dx} = \left(\frac{n\pi}{L}\right)
\left[\sqrt{\frac{2}{L}} \cos\left(\frac{n\pi x}{L}\right)\right]
\qquad
(0 \leq x \leq L),
\end{equation}
and
\begin{equation}
\frac{d^2\psi_n(x)}{dx^2} = -\left(\frac{n\pi}{L}\right)^2
\left[\sqrt{\frac{2}{L}} \sin\left(\frac{n\pi x}{L}\right)\right]
+ \frac{n\pi}{L}\sqrt{\frac{2}{L}}
\left[\delta(x) + (-1)^{n+1}\delta(x-L)\right] .
\end{equation}
Using either form of Eq.~(\ref{x_ke}), we find that
\begin{equation}
\langle \psi_n |\hat{T}|\psi_n \rangle
= \frac{1}{2m}
\langle \psi_n |\hat{p}^2 |\psi_n\rangle
= \frac{\hbar^2 n^2\pi^2}{2mL^2}
= \frac{p_n^2}{2m}
\end{equation}
where $p_n \equiv n\pi \hbar/L$. We also note that expectation
values of higher order powers of $\hat{p}$ are ill-defined.

The momentum-space wavefunction for the infinite well is given by
\begin{subequations}
\begin{align}
\phi_n(p) & = \frac{1}{\sqrt{2\pi \hbar}}
\int_{0}^{L} \psi_n(x)\,e^{-ipx/\hbar}\, dx \\
& = \sqrt{\frac{\hbar}{2\pi}}
\,
\left(\sqrt{\frac{2}{L}}\right)
\left[ (-1)^n\,e^{-ipL/\hbar} -1\right] \frac{p_n}{(p^2 - p_n^2)}
\label{isw_momentum_space}.
\end{align}
\end{subequations}
The expectation value of $p^2$ can be evaluated via
\begin{equation}
\langle \phi_n| p^2 | \phi_n \rangle
= \!\int_{-\infty}^{+\infty} p^2\,|\phi_n(p)|^2\,dp
= p_n^2
\end{equation}
using standard integrals, symbolic manipulation software, or by
contour integration techniques. Here again, higher powers of $p$ are not
well-defined expectation values because $|\phi(p)| \sim 1/p^2$. We also
see oscillating behavior as expected from the interference of contributions
from two singular potentials (the two walls).

The large $|p|$ behavior of
Eq.~(\ref{isw_momentum_space}) is consistent
with the simple form in Eq.~(\ref{first_general_result}),
using the contribution from
$x=0$ and $x=L$ and the values
\begin{align}
\psi_n'(0^{-}) & = \psi_n'(L^{+}) = 0 \\
\psi_n'(0^{+}) & = \sqrt{\frac{2}{L}}\, \left(\frac{n\pi}{L}\right) \\
\psi_n'(L^{-}) & = \cos(n\pi)
\sqrt{\frac{2}{L}}\, \left(\frac{n\pi}{L}\right)
=
(-1)^{n}
\sqrt{\frac{2}{L}}\, \left(\frac{n\pi}{L}\right).
\end{align}

The form in Eq.~(\ref{first_general_result}) has been found
in the context of two simple systems, ones with
singular potentials added to otherwise free particle systems. We next turn to a nontrivial system to which an
impenetrable boundary has been added, to further explore the generality of the
result in Eq.~(\ref{first_general_result}).

\section{Quantum bouncer}
\label{sec:quantum_bouncer}

Another pedagogically familiar system which includes a single infinite wall
barrier is the quantum bouncer,\cite{bouncer_1} defined by the potential
\begin{equation}
V(z) =
\begin{cases}
Fz & (z\geq 0) \\
\infty & (z<0)
\end{cases}.
\end{equation}
This potential has received renewed interest as the simplest
model for recent experiments showing evidence for quantized energy
states of neutrons in the Earth's gravitational field.\cite{neutron_bound_states}
Other recent applications of this model
to physical problems are discussed in Refs.~\onlinecite{app_1,app_2,app_3}.

The properties of the Airy function solutions have been re-examined
in the light of this renewed interest with new analytic results,\cite{bouncing_ball,vallee,goodmanson,airy_sum_rules,stark_linear_potentials}
using identities which appeared some time ago.\cite{albright}
For example, the normalized wavefunctions are,
\begin{equation}
\psi_n(z) =
\begin{cases}
N_n\, Ai(y-\zeta_n) & (y>0) \\
0 & (y\leq 0)
\end{cases}
\label{bouncer_wavefunction}
\end{equation}
\mbox{where}
\begin{equation}
N_n = \frac{1}{\sqrt{\rho} Ai'(-\zeta_n)}.
\end{equation}
The relevant dimensionless quantities are
\begin{equation}
y = \frac{z}{\rho}
\quad
\mbox{and}
\quad
\rho = \left(\frac{\hbar^2}{2mF}\right)^{1/3}.
\label{bouncer_properties_1}
\end{equation}
The $-\zeta_n$ are the zeros of the well-behaved
Airy function, $Ai(z)$,
and the energies are given in terms of them as
\begin{equation}
E_n = {\cal E}_0 \zeta_n
\quad
\mbox{where}
\quad
{\cal E}_0 \equiv F\rho = \left(\frac{\hbar^2F^2}{2m}\right)^{1/3}.
\label{bouncer_properties_2}
\end{equation}
The corresponding classical position-space probability density is
\begin{equation}
P^{(\rm CL)}_n(x) = \frac{1}{2\sqrt{A_n(A_n-z)}}
\qquad
\mbox{($0 \leq z \leq A_n$)}
\end{equation}
and zero elsewhere. The upper classical turning point is given by $\zeta_n {\cal E}_0 = E_n = FA_n$, or
$A_n = \rho \zeta_n$.
The classical momentum-space probability density is given by
\begin{equation}
P^{(\rm CL)}_n(p) =
\begin{cases}
\dfrac{1}{2Q_n} & (-Q_n \leq p \leq +Q_n) \\
0 & \mbox{otherwise}
\end{cases}
\label{classical_bouncer_momentum_one}
\end{equation}
where
\begin{equation}
\zeta_n{\cal E}_0 = E_n = \frac{Q_n^2}{2m}
\quad
\mbox{or}
\quad
Q_n = \frac{\hbar}{\rho} \sqrt{\zeta_n} .
\end{equation}
We note that the classical and quantum probability densities
for both position and
momentum for a closely related system, the symmetric linear
potential\cite{robinett_classical_quantum}
(discussed in Sec.~\ref{sec:symmetric_linear_potential}),
have been compared.

The momentum-space wavefunctions can be obtained
numerically by using the $\psi_n(z)$ in Eq.~(\ref{bouncer_wavefunction})
and the definition of the Fourier transform,
\begin{subequations}
\begin{align}
\tilde{\phi}_n(p) & = \frac{1}{\sqrt{2\pi \hbar}}
\int_{0}^{+\infty} \tilde{\psi}_n(z)\, e^{-ipz/\hbar}\,dz \\
& = \frac{1}{\sqrt{2\pi \hbar}}\!
\int_{0}^{+\infty}
\psi_{n}(z)\, \left[
\cos\left(\frac{pz}{\hbar} \right)
- i
\sin\left(\frac{pz}{\hbar} \right) \right]\, dz \\
& \equiv \tilde{\phi}_n^{(\rm RE)} (p) - i\tilde{\phi}_n^{(\rm IM)} (p).
\label{quantum_bouncer_fourier_transform}
\end{align}
\end{subequations}
We plot the contributions
of $|\tilde{\phi}_n^{(\rm IM)}(p)|^2$, $|\tilde{\phi}_n^{(\rm RE)}(p)|^2$, and
their sum, each compared to $P_n^{(\rm CL)}(p)$, in
Figs.~\ref{fig:quantum_bouncer}(a), (b), and (c). 

Note that the contribution of the imaginary component of $\tilde{\phi}(p)$
shown in Fig.~1(a)
is similar to many of the visualizations of comparisons between
classical and quantum mechanical probability distributions, where the
quantum result oscillates about the classical prediction, consistent
with WKB-type approximations. In this case,
the imaginary component provides half of the total probability, in that
``locally averaged'' sense. The real part [see Fig.~1(b)] has similar
oscillatory behavior, but with a slightly different structure. The
combination of the two gives a much smoother approach to the 
classical ``flat'' momentum distribution.

More importantly, we see that the real component of the Fourier transform
(the one affected most directly by the infinite wall)
extends much further into the classical disallowed region of momentum space,
hinting at the expected power-law `tail'.
We can fit the large $|p|$ tails of $\tilde{\phi}^{(\rm RE)}_{n}(p)$ for various values of
$n$ and find the simple result ($\rho$ and
$\hbar$ set equal to unity)
\begin{equation}
p^2 |\tilde{\phi}_n^{(\rm RE)}(p)| \approx 0.4
\quad
\mbox{($p\rightarrow \infty$)} ,
\end{equation}
independent of $n$. To compare this result to the prediction of
Eq.~(\ref{first_general_result}), we note that $\psi'_n(0^{-}) = 0$,
and from the normalized wavefunction in Eq.~(\ref{bouncer_wavefunction}),
we find that
\begin{equation}
\psi'_n(0^{+}) = \frac{1}{\sqrt{\rho} Ai'(-\zeta_n)}
Ai'_n(-\zeta_n) = \frac{1}{\sqrt{\rho}},
\end{equation}
which is independent of $n$, so that in dimensionless form
\begin{equation}
p^2 |\tilde{\phi}_n^{(RE)}(p)| \sim \frac{1}{\sqrt{2\pi}}
\approx 0.3989,
\end{equation}
which is another important confirmation of the simple result in
Eqn.~(\ref{first_general_result}) for singular potentials. 
(We note that a similar numerical
evaluation and subsequent fitting of $\tilde{\phi}_{n}^{(\rm IM)}(p)$
finds that it scales as $1/p^5$. We will discuss this result in
Sec.~\ref{sec:symmetric_linear_potential}.)

\section{\label{sec:hints}Hybrid example: Hints of the general solution}

We next consider potentials that are discontinuous, but not singular. The most familiar example is a step potential, defined
by
\begin{equation}
V_{s}(x;a) = V_0\Theta (x-a)
\label{single_step_potential}
\end{equation}
where $\Theta(x)$ is the Heaviside function. For a finite well (FW) we have
can write
\begin{equation}
V_{\rm FW}(x;a,b) = -V_{s}(x;a) + V_{s}(x;b)
=
\begin{cases}
0 & \mbox{($x<a$)} \\
-V_0 & \mbox{($a<x<b$)} \\
0 & \mbox{($x>b$)}
\end{cases}
\label{finite_well_potential} .
\end{equation}
The boundary conditions for such
discontinuous potentials are that both $\psi$ and $\psi'$ are
continuous across such a step,\cite{branson} and thus we might expect a qualitatively different
behavior of $\phi(p)$, due to the poor behavior of $\psi''$ at
a step boundary.

To pursue this question and to allow for a more
systematic study of the large $|p|$ behavior of $\phi(p)$,
we consider the hybrid
case of an attractive $\delta$-function potential combined with a
single step potential. For definiteness, we consider
\begin{equation}
V(x) = V_{\delta}(x;0) + V_{s}(x;a),
\end{equation}
which is a slight generalization of a single $\delta$-function
potential interacting with an infinite wall, as discussed
by Aslangul.\cite{delta_plus_wall}
We note that for sufficiently large positive $V_0$, the single bound state
is no longer supported, and if $V_0<0$, the possibility of tunneling
can also preclude a stable bound state. Although the study of what range of
$V_0$ values support a bound state is an interesting
question in itself, we assume that $V_0$ is such that there is one,
with energy $E = -|E|$, and focus on the behavior of the
corresponding $\phi(p)$.
We do not need to find a normalized solution in detail because
we will focus only on the nature of the wavefunction (and any discontinuities)
and the locations of the singularity and discontinuity in the potential.

The (un-normalized) solutions for each region can be written as
\begin{equation}
\psi(x) =
\begin{cases}
Ae^{+Kx} & \mbox{($x<0$)} \\
Be^{-Kx} + Ce^{+Kx} & \mbox{($0<x<a$)} \\
De^{-Qx} & \mbox{($x>a$)}
\end{cases},
\end{equation}
where
\begin{equation}
K^2 = \frac{2m|E|}{\hbar^2}
\quad
\mbox{and}
\quad
Q^2 = \frac{2m(|E|+V_0)}{\hbar^2} .
\end{equation}
The boundary conditions which are imposed by the singular $\delta$-function
and discontinuous step potential are
\begin{subequations}
\begin{align}
&\psi(x=0) && A = B + C
\label{cont1} \\
&\psi(x=a) && Be^{-Ka} + C^{+Ka} = De^{-Qa}
\label{cont2}\\
&\psi'(x=a) && -KBe^{-Ka} + KCe^{+Ka} = -QDe^{-Qa}
\label{cont3} .
\end{align}
\end{subequations}
The corresponding Fourier transform is
\begin{subequations}
\begin{eqnarray}
\phi(p) & = & \frac{1}{\sqrt{2\pi \hbar}}
\left\{
\frac{A}{K - ip/\hbar} + \frac{B}{K+ip/\hbar} - \frac{C}{K -ip/\hbar}
\right\} \nonumber \\
&&{}
+ \frac{1}{\sqrt{2\pi\hbar}}
\left\{
-\frac{Be^{-Ka}}{K +ip/\hbar} + \frac{Ce^{+Ka}}{K -ip/\hbar}
+ \frac{De^{-Qa}}{Q + ip/\hbar}
\right\}
e^{-ipa/\hbar} \\
& = &
\phi_{\delta}(p) + \phi_{s}(p),
\end{eqnarray}
\end{subequations}
where we have separated the terms related to the singularity and discontinuity.

Because we are interested in the large $|p|$ behavior, we can systematically
expand each term in inverse powers of $p$. For example, we find that
for the $\delta$-function contribution
\begin{equation}
\phi_{\delta}(p)
= \frac{1}{\sqrt{2\pi\hbar}}
\left[
-i (A-B-C)\frac{\hbar}{p}
+ (AK +BK -CK)\frac{\hbar^2}{p^2}
+ \cdots
\right].
\label{phi_delta}
\end{equation}
The first term vanishes because of the continuity of $\psi$ at the
origin, and the second one is consistent with the general form for
singular potentials in Eq.~(\ref{first_general_result}) (with the
singularity located at $x=0$ so that $\exp(-ipa/\hbar) = 1$.)

For the term arising from the single-step potential, we have
\begin{eqnarray}
\phi_{s}(p)
& = & \frac{1}{\sqrt{2\pi\hbar}}
\Big[
i (Be^{-Ka} + Ce^{+Ka} - De^{-Qa})\frac{\hbar}{p}
+ (-Bke^{-Ka} + CKe^{+Ka} + Dqe^{-Qa})\frac{\hbar^2}{p^2}
\nonumber \\
& & {}
+ i (-BK^2e^{-Ka} -CK^2e^{+Ka} + DQ^2e^{-Qa})\frac{\hbar^3}{p^3}
+\cdots
\Big] e^{-ipa/\hbar},
\label{phi_wall}
\end{eqnarray}
where we have carried the expansion to one higher order. The first two terms
vanish because of the continuity of $\psi$ and $\psi'$
given by Eqs.~(\ref{cont1}) and (\ref{cont2}) respectively. The remaining non-vanishing higher order term can be expressed as
\begin{equation}
\frac{i}{\sqrt{2\pi \hbar}}
\left[\psi''(a^{+}) - \psi''(a^{-})\right]\,e^{-ipa/\hbar}
\frac{\hbar^3}{p^3}.
\label{next_nontrivial}
\end{equation}
Note that for discontinuous potentials, we can derive a relation
similar to Eq.~(\ref{cusp_condition}) for a step-potential of the
form in Eq.~(\ref{single_step_potential}). By differentiating the
Schr\"odinger equation once, and then integrating over the
range $(a^{-},a^{+})$ we find that
\begin{equation}
\psi''(a^{+}) - \psi''(a^{-}) = \frac{2mV_0}{\hbar^2}\, \psi(a)
\label{next_cusp_condition}
\end{equation}
because
$V'_{s}(x;a) = V_0\delta(x-a)$, which can be used to simplify Eq.~(\ref{next_nontrivial}).

Motivated by these results, we have confirmed that the 
expression in Eq.~(\ref{next_nontrivial})
gives the leading large $|p|$ behavior
of $\phi(p)$ for a variety of
model systems involving discontinuous step potentials,
including the finite well of Eq.~(\ref{finite_well_potential}),
the asymmetric finite well (with different barrier heights on either side),
and other combinations.

More importantly, if we use this problem as a template,
we find that we can express the series expansion for each term,
$\phi_{\delta,s}(p)$, in Eqs.~(\ref{phi_delta}) and (\ref{phi_wall}),
in the common form
\begin{equation}
\phi(p) = \frac{1}{\sqrt{2\pi \hbar}}
\sum_{k=0}^{\infty} {\cal T}_k(p),
\label{general_expansion}
\end{equation}
where
\begin{subequations}
\label{form_all}
\begin{align}
{\cal T}_{1}(p) & = -i \left[\psi(a^{+}) - \psi(a^{-})\right]\,e^{-ipa/\hbar}\, \frac{\hbar}{p}
\label{form_1}\\
{\cal T}_{2}(p) & = - \left[\psi'(a^{+}) - \psi'(a^{-})\right]\,e^{-ipa/\hbar}\, \frac{\hbar^2}{p^2} \label{form_2}\\
{\cal T}_{3}(p) & = +i \left[\psi''(a^{+}) - \psi''(a^{-})\right]\,e^{-ipa/\hbar}\, \frac{\hbar^3}{p^3} \label{form_3}\\
{\cal T}_{4}(p) & = + \left[\psi'''(a^{+}) - \psi'''(a^{-})\right]\,e^{-ipa/\hbar}\, \frac{\hbar^4}{p^4} \label{form_4}\\
\vdots \nonumber \\
{\cal T}_{n}(p) & = (-i)^n [\psi^{(n-1)}(a^{+}) - \psi^{(n-1)}(a^{-})]\, e^{-ipa/\hbar}
\left(\frac{\hbar}{p}\right)^{n}.
\label{form_n}
\end{align}
\end{subequations}

In each case, the vanishing of the leading ${\cal O}(p^{-1})$ term is a
consequence of $\psi(x)$ being everywhere
continuous, which is required for a consistent probability interpretation.
This condition then implies that
the lowest-order term possible for $\phi(p)$
is of order ${\cal O}(p^{-2}$) for large $|p|$,
which ensures that $\langle p^2\rangle$
always gives a convergent integral.

\section{Formal solution}
\label{sec:formal_solution}

To confirm the systematic expansion of $\phi(p)$
suggested by Eq.~(\ref{form_all}),
we now describe an approach to the evaluation of the
Fourier transform $\phi(p)$, focusing on the case where the
corresponding $\psi(x)$ has a discontinuous derivative at some
order. For simplicity, we assume that the generalized kink is at $x=0$; the extension to any other location is trivial.

We first write the Fourier transform as
\begin{subequations}
\begin{align}
\phi(p) & = \frac{1}{\sqrt{2\pi \hbar}} \int_{-\infty}^{+\infty}
\psi(x)\, e^{-ipx/\hbar}\,dx \\
& = \frac{1}{\sqrt{2\pi \hbar}} \int_{-\infty}^{+\infty}
\psi(x)\, \cos\left(\frac{px}{\hbar}\right) \, dx
- i
\frac{1}{\sqrt{2\pi \hbar}} \int_{-\infty}^{+\infty}
\psi(x)\, \sin\left(\frac{px}{\hbar}\right) \, dx \\
& \equiv \frac{1}{\sqrt{2\pi \hbar}}\left[I_1(p) - i I_2(p)\right],
\end{align}
\end{subequations}
and we will focus on the approximation of $I_{1,2}(p)$ for large $|p|$.

Assuming that any continuity issues are localized at $x=0$, we split
the integral into two regions, adding appropriate convergence
factors, $\exp(\pm \epsilon x)$, namely.
\begin{subequations}
\begin{align}
I_1(p) & = \int_{-\infty}^{+\infty}
\psi(x)\, \cos\left(\frac{px}{\hbar}\right) \, dx \\
& = \lim_{\epsilon \rightarrow 0}
\left[
\int_{-\infty}^{0} \psi(x)\, \cos\left(\frac{px}{\hbar}\right) \, e^{+\epsilon x}\, dx
+
\int_{0}^{+\infty} \psi(x)\, \cos\left(\frac{px}{\hbar}\right) \, e^{-\epsilon x}\, dx \right] \\
& \equiv \lim_{\epsilon \rightarrow 0}
\left[I_1^{(\epsilon)}(p)\right].
\end{align}
\end{subequations}
We can first rewrite the cosine term and expand $\psi(x)$ in a series
expansion (one for each integration region) via
\begin{align}
\cos\left(\frac{px}{\hbar}\right) &= \frac{1}{2}\left(e^{ipx/\hbar}
+ e^{-ipx/\hbar}\right), \\
\noalign{\noindent and}
\psi(x) &= \sum_{n=0}^{\infty} \frac{\psi^{(n)}(0)x^n}{n!},
\end{align}
giving
\begin{eqnarray}
I_1^{(\epsilon)}(p) & =&
\frac{1}{2}
\left\{
\sum_{n=0}^{\infty}
\frac{\psi^{(n)}(0^{-})}{n!}
\int_{-\infty}^{0} x^n \left[
e^{x(\epsilon + ip/\hbar)}
+
e^{x(\epsilon - ip/\hbar)}
\right] \, dx\right.\nonumber \\
& &{}
\left.
+
\sum_{n=0}^{\infty}
\frac{\psi^{(n)}(0^{+})}{n!}
\int_{0}^{+\infty} x^n \left[
e^{-x(\epsilon - ip/\hbar)}
+
e^{-x(\epsilon + ip/\hbar)}
\right] \,dx \right\}.
\end{eqnarray}
We use $0^{-}$ ($0^+$) for the $x<0$ ($x>0$) integrals respectively.
(This type of regularization method is similar to that used in
establishing the completeness relations of the eigenfunctions of the
single $\delta$-function potential.\cite{completeness})

Because of the convergence factors, the integrals are easily performed using
\begin{equation}
\int_{-\infty}^{0}\, y^n\,e^{y}\,dy = (-1)^n n!
\quad
\mbox{and}
\quad
\int_{0}^{+\infty}\, y^n\,e^{-y}\,dy = n!.
\label{needed_integrals}
\end{equation}
These integrals give
\begin{subequations}
\begin{align}
I_1^{(\epsilon)}(p) & = \left(\frac{1}{2}\right)
\sum_{n=0}^{\infty}
\left[
\frac{1}{(\epsilon - ip/\hbar)^{n+1}}
+
\frac{1}{(\epsilon + ip/\hbar)^{n+1}}
\right]
\left[
\psi^{(n)}(0^{+}) + (-1)^n\psi^{(n)}(0^{-})
\right] \\
& \rightarrow \left(\frac{1}{2}\right)
\sum_{n=0}^{\infty}
\left[
\frac{i^{n+1}[1+(-1)^{n+1}]\hbar^{n+1}}{p^{n+1}}
\right]
\left[\psi^{(n)}(0^{+}) + (-1)^n\psi^{(n)}(0^{-})\right]
\quad
\mbox{as $\epsilon \rightarrow 0$}\\
& = - \frac{\hbar^2}{p^2}[\psi'(0^{+}) - \psi'(0^{-})]
+ \frac{\hbar^4}{p^4}[\psi'''(0^{+}) - \psi'''(0^{-})]
- \frac{\hbar^6}{p^6}[\psi^{(5)}(0^{+}) - \psi^{(5)}(0^{-})]
+ \cdots .
\end{align}
\end{subequations}
Thus, the cosine component of the Fourier transform giving $\phi(p)$
gives the systematic expansion in differences of odd powers of
derivatives ($\psi^{(2k+1)}(0)$) at the discontinuity,
just as in Eqs.~(\ref{form_2}) and (\ref{form_4}).

In the same manner, we can evaluate the $\sin(px/\hbar)$ integral
and find
\begin{subequations}
\begin{align}
-iI_{2}^{(\epsilon)}(p) & = \left(\frac{-i}{2i}\right)
\sum_{n=0}^{\infty}
\left[
\frac{1}{(\epsilon - ip/\hbar)^{n+1}}
-
\frac{1}{(\epsilon + ip/\hbar)^{n+1}}
\right]
[
\psi^{(n)}(0^{+}) - (-1)^n\psi^{(n)})(0^{-})
] \\
& \rightarrow \left(-\frac{1}{2}\right)
\sum_{n=0}^{\infty}
\left[
\frac{i^{n+1}[1-(-1)^{n+1}]\hbar^{n+1}}{p^{n+1}}
\right]
[\psi^{(n)}(0^{+}) + (-1)^n\psi^{(n)}(0^{-})]
\quad
\mbox{as $\epsilon \rightarrow 0$} \\
& = - i\frac{\hbar}{p}[\psi(0^{+}) - \psi(0^{-})]
+ i\frac{\hbar^3}{p^3}[\psi''(0^{+}) - \psi''(0^{-})]
- i\frac{\hbar^5}{p^5}[\psi^{(4)}(0^{+}) - \psi^{(4)}(0^{-})]
+ \cdots,
\end{align}
\end{subequations}
which gives the appropriate even powers of derivatives, along
with the correct factors of $\pm i$, reproducing the results
in Eqs.~(\ref{form_1}) and (\ref{form_3}). Taken together, these two results
give the general form in Eq.~(\ref{form_n}).

If the discontinuity is located at another location, we can use the result
that if $\psi(x) \rightarrow \psi(x-a)$, then the corresponding momentum-space
wavefunction satisfies $\phi(p) \rightarrow \phi(p)\,\exp(-ipa/\hbar)$ with
the appropriate derivatives now evaluated at $a^{\pm}$.

If the potential (and resulting solutions) are well-behaved, as with the
harmonic oscillator, then all such differences of the derivatives vanish,
implying that there are no power-law tails. This connection is not surprising
because such an asymptotic expansion will not capture information on more
well-behaved (for example, $\exp(-\alpha^2 p^2/2)$) functions.

As noted, if the potential has a generalized
discontinuity,
then repeated differentiation of the Schr\"{o}dinger equation yields
a relation between the lowest-order difference in derivatives and the
wavefunction at the discontinuity. Examples include the relations in
Eqs.~(\ref{cusp_condition}) and (\ref{next_cusp_condition}). If $V^{(k)}(x)$ has a discontinuity at $x=a$, then $dV^{(k+1)}(x)/dx
\propto \delta(x-a)$ and differentiating the Schr\"{o}dinger equation $k+1$
times, and then integrating over the range $(a-\epsilon,a+\epsilon)$ yields
a relation of the form
\begin{equation}
\psi^{(k+2)}(a^{+}) - \psi^{(k+2)}(a^{-})
=
\frac{2m}{\hbar^2} \left[
V^{(k)}(a^{+})
-
V^{(k)}(a^{-})
\right] \psi(a),
\label{general_cusp_condition}
\end{equation}
which can be used to simplify the leading order term in
Eq.~(\ref{form_n}).
In the case where $\psi(a) = 0$ (where the wavefunction
vanishes at the discontinuity), the expansion will
necessarily start with one more power of $p^{-1}$. In that case, the
first non-vanishing term in the $(k+1)$th differentiation of the
$V(x)\psi(x)$ term gives
\begin{equation}
\psi^{(k+3)}(a^{+}) - \psi^{(k+3)}(a^{-})
=
(k+1) \frac{2m}{\hbar^2} \left[
V^{(k)}(a^{+})
-
V^{(k)}(a^{-})
\right] \psi'(a),
\label{higher_order_cusp_condition}
\end{equation}
which can be used in the ${\cal T}_{k+4}(p)$ term in the expansion
in Eq.~(\ref{general_expansion}). A case where this
situation occurs is considered in Sec.~\ref{sec:symmetric_linear_potential}.

\section{The symmetric linear potential}
\label{sec:symmetric_linear_potential}

We now consider another example
to confirm the higher order predictions, and to note one
additional novel feature. A straightforward generalization
of the quantum bouncer is the symmetric linear potential, defined by
\begin{equation}
V(z) = F|z|.
\end{equation}
which has been described as a
``parity extended'' version of the quantum bouncer\cite{parity_extended}
and shares many of the same features. For this potential, which has a
cusp (discontinuity in $V'(x)$), we expect that
$\phi(p) \propto 1/p^4$ as in Eq.~(\ref{form_4}),
at least for those states with $\psi(0) \neq 0$
as in Eq.~(\ref{general_cusp_condition}).
We note for future reference that $V''(z) = 2F\delta(z)$.

Because of the symmetric nature of this potential, the
eigenstates have definite parity. The odd states are simply related to
those of the quantum bouncer and using the same notation as in
Sec.~\ref{sec:quantum_bouncer} we have
\begin{equation}
\psi_n^{(-)}(z)
=
\frac{1}{\sqrt{2}}
\begin{cases}
+N_n\, Ai(y-\zeta_n) & \mbox{($y>0$)} \\
-N_n\, Ai(-y-\zeta_n) & \mbox{($y<0$)}
\end{cases},
\end{equation}
with the same notation and energies as in Eqs.~(\ref{bouncer_properties_1})
and (\ref{bouncer_properties_2}).
The corresponding (properly normalized) even states can be written in the form
\cite{parity_extended}
\begin{equation}
\psi_n^{(+)}(z) = M_n\,Ai(|y|-\eta_n)
\quad
\mbox{where}
\quad
M_n \equiv \frac{1}{\sqrt{2\rho\eta_n}\,Ai(-\eta_n)}.
\end{equation}
The $-\eta_n$ are the zeros of the derivative of the
well-behaved Airy function, given by $Ai'(-\eta_n) = 0$.

For this potential we expect the leading order term
in the large-$|p|$ expansion of $\phi(p)$ in Eq.~(\ref{form_n})
to be
\begin{equation}
\frac{1}{\sqrt{2\pi\hbar}}
\left[\psi'''(a^{+}) - \psi'''(a^{-})\right] e^{-ipa/\hbar}\,
\frac{\hbar^4}{p^4},
\end{equation}
which depends on the third derivatives of $\psi$.
If we use the same approach leading to Eqs.~(\ref{cusp_condition}) and
(\ref{next_cusp_condition}), and more generally
Eq.~(\ref{general_cusp_condition}), namely
integrating the Schr\"odinger equation (differentiated
an appropriate number of times) we find a constraint on $\psi'''$,
namely,
\begin{equation}
\frac{\hbar^2}{2m}\left[\psi'''(0^{+}) - \psi'''(0^{-})\right]
= 2F\psi(0),
\end{equation}
or
\begin{equation}
\left[\psi'''(0^{+}) - \psi'''(0^{-})\right]
= \frac{2\psi(0)}{\rho^3}.
\label{even_phi_limit}
\end{equation}
The momentum-space solutions corresponding to even states of
this potential, where $\psi(0) \neq 0$, are expected to scale as
\begin{equation}
p^4\, |\phi^{(+)}_{n}(p)|
\longrightarrow
\frac{1}{\sqrt{2\pi}} \frac{2}{\sqrt{2\eta_n}}
\label{even_symmetric_linear_case}
\end{equation}
when all dimensional quantities are removed. This result has an $n$-dependence
through $\eta_n$, which scales as $n^{2/3}$ for large values of $n$.
\cite{parity_extended} (Note that the $\phi^{(+)}_n(p)$ receives
contributions only from the $\cos(px/\hbar)$ term in the Fourier transform.)

In contrast, the odd states have a vanishing value of $\psi(0)$, and
thus necessarily have an asymptotic $|p|$ dependence for $\phi(p)$
that starts at one order higher, namely proportional to ${\cal T}_{n=5}
\sim p^{-5}$. For
this case we can also simplify the general expression by writing
\begin{equation}
\psi''''(0^{-}) - \psi''''(0^{+}) = - \frac{4\psi'(0)}{\rho^3},
\label{odd_phi_limit}
\end{equation}
which is most easily obtained by repeated differentiation of the Airy
differential equation on either side of the cusp at $x=0$, as in
Eq.~(\ref{higher_order_cusp_condition}). For comparison to numerically
obtained results, we once again set dimensional quantities equal to unity, 
in which case the large $|p|$ behavior of $\phi(p)$ is expected to be
\begin{equation}
p^5\, |\phi^{(-)}_{n}(p) |
\longrightarrow
\frac{1}{\sqrt{2\pi}}\frac{4}{\sqrt{2}}
= \frac{2}{\sqrt{\pi}},
\label{odd_symmetric_linear_case}
\end{equation}
independent of $n$. (In this case $\phi_{n}^{(-)}(p)$ receives
contributions only from the $\sin(px/\hbar)$ term in the Fourier transform.)
We have numerically evaluated $\phi^{(\pm)}_n(p)$ for several values of
$n$ and present the results for representative even and odd cases ($n=11$) in
Fig.~\ref{fig:linear_tail}, along with the large $|p|$ approximations
in Eqs.~(\ref{even_symmetric_linear_case})
and (\ref{odd_symmetric_linear_case})
and find excellent agreement.

Note that the $1/p^5$ behavior of the imaginary component
of $\tilde{\phi}_n(p)$ for the quantum bouncer, which here is
directly related to $\phi_{n}^{(-)}(p)$, can be understood as a
special case of the general results
in Eq.~(\ref{higher_order_cusp_condition}).

\section{Conclusions}

The ability to make connections between the potential of a system
and the resulting wavefunctions in position space is an important part of the toolkit of
any quantum mechanic. Being able to visualize these connections and
understand them at a conceptual level is an increasingly important focus
of pedagogy \cite{theme_issue} in the subject. 

\begin{figure}[hbt]
\epsfig{file=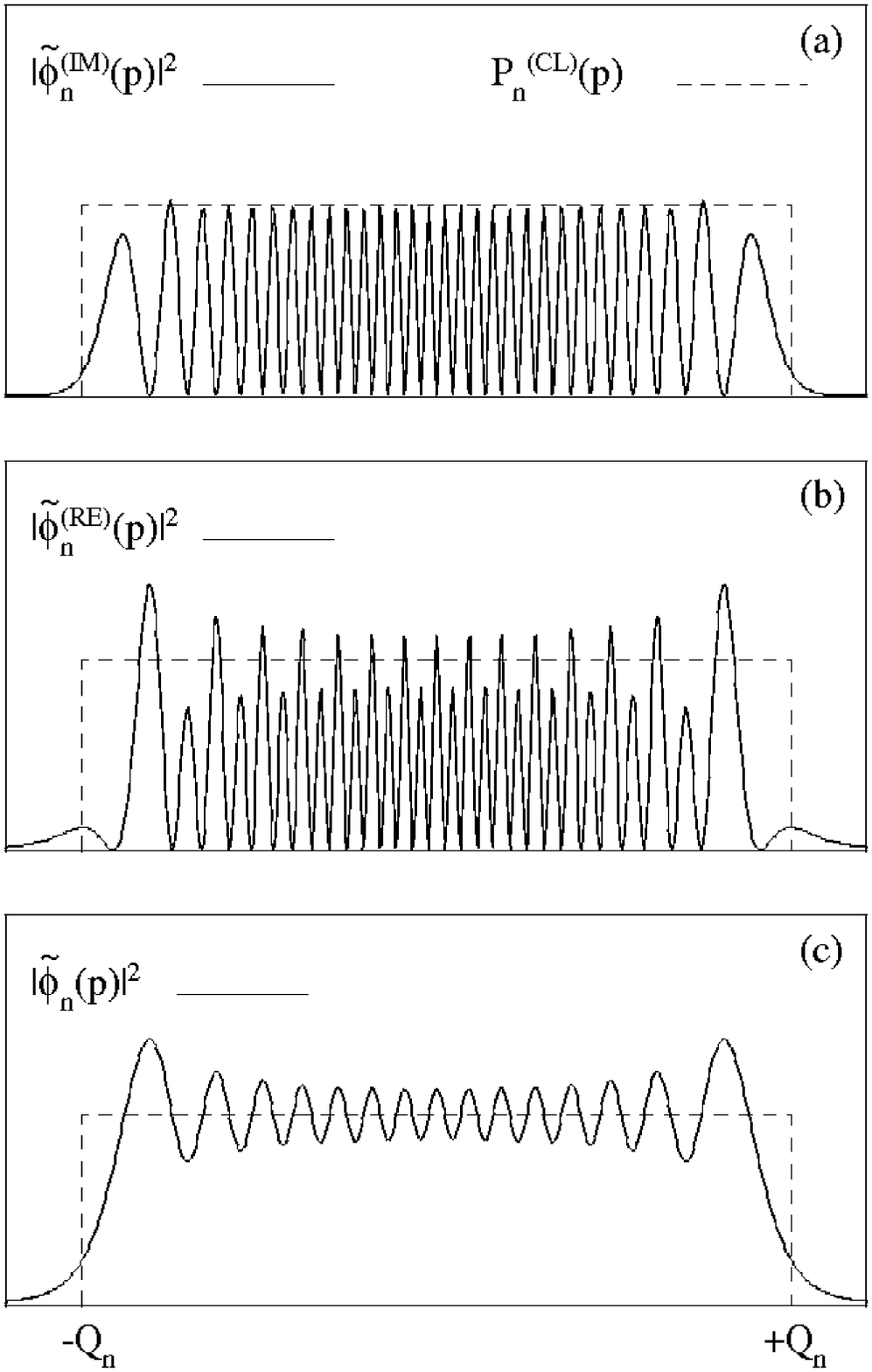,width=0.6\linewidth,angle=0}
\caption{
Plots of the classical momentum-space probability distribution,
$P_{n}^{(CL)}(p)$ versus $p$,
for the quantum bouncer (dashed curve)
from Eq.~(\ref{classical_bouncer_momentum_one}) and
the corresponding quantum distribution, $|\tilde{\phi}(p)|^2$ (solid curves)
from Eq.~(\ref{quantum_bouncer_fourier_transform}).
The imaginary and real parts of $|\tilde{\phi}(p)|^2$ from
are shown
in (a) and (b) and the total $|\tilde{\phi}(p)|^2$ in (c).
The vertical dashed lines labeled $\pm Q_n$ indicate
the classical turning points in momentum-space.}
\label{fig:quantum_bouncer}
\end{figure}

\begin{figure}[hbt]
\epsfig{file=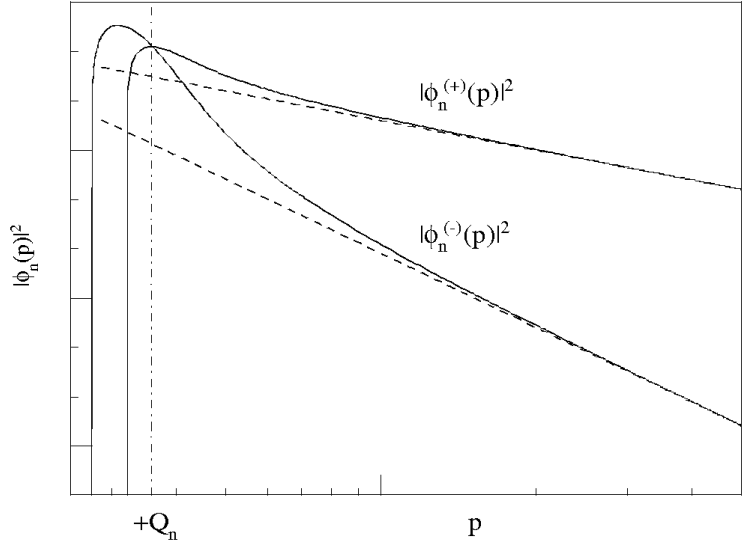,width=0.5\linewidth,angle=270}
\caption{The solid curves show $|\phi(p)|^2$ for the
symmetric linear potential in Sec.~\ref{sec:symmetric_linear_potential}
obtained by numerical Fourier transform for a typical even
($\phi_n^{(+)}(p)$) and odd ($\phi_n^{(-)}(p)$) case ($n=11$).
The straight dashed lines
correspond to the lowest-order non-vanishing predictions in
Eqs.~(\ref{even_symmetric_linear_case})
and (\ref{odd_symmetric_linear_case}). The vertical dot-dash line
indicates the classical turning point in momentum-space at
$+Q_n$ for positive $p$. Note the log-log scales.}
\label{fig:linear_tail}
\end{figure}

The corresponding skills involved
in understanding and interpreting quantum phenomena
in momentum space are far less developed. Some simple
examples in this area exist, such as the ``flat'' momentum distribution for
the quantum bouncer (as in Fig.~\ref{fig:quantum_bouncer}) arising from
the constant force law for that system, and some others.
\cite{robinett_classical_quantum,robinett_momentum} Any
new examples that provide tangible relations between the potential energy
function and $\phi(p)$ in a direct way are valuable.
Our results
extend the study of the momentum-space probability
distribution from the semi-classical limit to the deeply quantum
regime of large momenta, far beyond the classical turning points
in $p$-space.

The single $\delta$-potential led us to some
early intuition about the structure of the general result, namely the form
in Eq.~(\ref{first_general_result}). This observation  was important as
it emphasized the likely appearance of differences of derivatives at
various orders depending on the behavior of $V(x)$.
As importantly, in generalizing this expression
to Eq.~(\ref{form_n}),
we noted that such a result would be consistent with
simple dimensional
analysis arguments, because for every higher derivative used, another power of
$\hbar/p$ would be required to compensate.

We suggest several possible
avenues for research amenable to exploration by motivated
undergraduates.
The system in Sec.~\ref{sec:symmetric_linear_potential} can, for example,
be extended to an asymmetric linear potential by having different
force constants for $x>0$ ($F$) and $x<0$ ($\overline{F}$).
The solutions can still be written
in terms of Airy functions, but the parity symmetry is now broken and all states should have $\phi(p)$ at large $|p|$ proportional to
$p^{-4}$, but with coefficients that depend on $(F-\overline{F})$. We can also imagine extending our work in other ways,
for example, by looking at if the
correlations found here between kinks in position-space
and tails in momentum-space are obvious in the Wigner distribution,
which is one of the canonical quantum mechanical formats for
discussing $x$-$p$ connections and correlations.

Extensions to more realistic three-dimensional systems are
also natural, such as finite well models of the nuclear force. The Coulomb problem for the hydrogen atom,
involves a singular potential, and the ground
state momentum-space wavefunction
\begin{equation}
\phi_{1S}(p) = \sqrt{\frac{8p_0^5}{\pi^2}}
\,
\frac{1}{(p^2+p_0^2)^2}
\end{equation}
(where $p_0 = \hbar/a_0$ with $a_0$ the Bohr radius)
gives only a finite number of well-defined expectation values for
$\langle p^k\rangle$ because of its power-law ($p^{-4}$)
behavior for large $|p|$.
Such momentum-space solutions are of interest
because $\phi_{1S}(p)$ has been directly measured
in scattering experiments. \cite{h_atom_momentum}

\noindent
\hfill

\hfill

\newpage

\noindent
\hfill

\hfill

\end{document}